# High-resolution Radio Study of Pulsar Wind Nebula MSH 15–52 and Supernova Remnant RCW 89

S. Zhang[1], C.-Y. Ng[1], and N. Bucciantini[2,3,4]
[1] Department of physics, The University of Hong Kong, Pokfulam Rd, HKSAR, Hong Kong, People's Republic of China; shumeng_zhang@connect.hku.hk
[2] INAF—Osservatorio Astrofisico di Arcetri, Largo Enrico Fermi 5, 50125, Firenze, Italy
[3] Dipartimento di Fisica e Astronomia, Università degli Studi di Firenze, Via Sansone 1, 50019, Sesto Fiorentino, Italy
[4] Istituto Nazionale di Fisica Nucleare, Sezione di Firenze, Via Sansone 1, 50019, Sesto Fiorentino, Italy

## Abstract

We present high-resolution radio observations of the pulsar wind nebula (PWN) MSH 15–52, which is renowned for its distinctive handlike shape, and its associated supernova remnant RCW 89. Using the Australia Telescope Compact Array, we obtained 3 and 6 cm radio maps with a resolution of 2″. These unveil small-scale radio features in the system and allow a direct comparison with the arcsecond-resolution X-ray images. We find that the radio emission is composed of a complex filamentary structure. In particular, there is a bar-like feature across the central pulsar B1509−58 in the inner PWN, and the radio sheath wrapping around the pulsar also appears to be made up of filaments. Some prominent X-ray features are not detected in radio, including the one-sided jet in the south and the fingerlike structures in the north. These indicate turnover of the particle distribution at low energies in these regions. For RCW 89, the radio emission well coincides with both the X-ray knots and the Hα filaments. The high polarization fraction shows that the emission is synchrotron in nature, but it extends well beyond the sharp boundary of the nonthermal X-ray emission, which is difficult to explain.

*Unified Astronomy Thesaurus concepts:* Pulsar wind nebulae (2215); Supernova remnants (1667); Polarimetry (1278)

## 1. Introduction

As pulsars spin down, most of their rotational energy is converted into a relativistic outflow, which forms pulsar wind nebulae (PWNe) upon interaction with the ambient medium (see B. M. Gaensler & P. O. Slane 2006 for a review). PWNe are important objects for understanding relativistic shock physics and the production of cosmic rays in our Galaxy. For young systems, they are generally bright and often exhibit an axisymmetric structure, including jets and arcs/torus, about the pulsar spin axis (see C. Y. Ng & R. W. Romani 2004, 2008). They provide the best cases for the study of pulsar interaction with the surrounding supernova ejecta and the coevolution of the nebula and the parent supernova remnant (SNR).

There are only a handful of PWNe younger than a few kiloyears, and one of them is MSH 15–52 (G320.4−1.2, Kes 23). This remarkable source has an age of 1.7 kyr (B. M. Gaensler et al. 1999) and is powered by the high magnetic field pulsar B1509−58 (J1513−5908), which has a spin-down luminosity of $2 \times 10^{37}$ erg s$^{-1}$ and surface $B$ field of $1.54 \times 10^{13}$ G. The PWN was first discovered in the radio band (B. Y. Mills et al. 1961; M. J. L. Kesteven 1968) and then subsequently in X-rays (F. D. Seward & F. R. Harnden 1982; F. D. Seward et al. 1983; C. Greiveldinger et al. 1995; K. T. S. Brazier & W. Becker 1997; D. Marsden et al. 1997; G. Cusumano et al. 2001) and gamma rays (T. Sako et al. 2000; F. Aharonian et al. 2005; M. Forot et al. 2006). High-resolution X-ray imaging with the Chandra X-ray Observatory reveals a complex structure of the PWN and the SNR RCW 89 in the north. The PWN X-ray emission consists of a pair of arcs wrapping around the pulsar, a one-sided jetlike outflow extending from the pulsar to southeast, and several fingerlike filaments pointing north (see Figure 1; B. M. Gaensler et al. 2002). The overall morphology resembles a hand; therefore, this system is known as the "cosmic hand." At the source distance of ∼5 kpc (B. M. Gaensler et al. 1999), MSH 15–52 has a physical size ≳32 pc, much larger than other young PWNe (e.g., the famous Crab Nebula is only ∼4 pc in size). Recently, IXPE made the first X-ray polarization measurement, and the result shows good $B$-field alignment with the jet and arc structure (R. W. Romani et al. 2023).

North of the PWN, the optical nebula RCW 89 (A. W. Rodgers et al. 1960) was suggested to be an associated SNR (B. M. Gaensler et al. 1999). Its X-ray emission consists of several bright compact knots arranged in a horseshoe shape (see Figure 1). It was proposed that these are hot spots resulting from the pulsar polar outflow interaction with the dense environment (e.g., K. T. S. Brazier & W. Becker 1997; Y. Yatsu et al. 2005). However, K. J. Borkowski et al. (2020) recently reported that the knots are composed of heavy elements and are moving outward at a high speed; they are, therefore, more likely to be supernova ejecta.

In the radio band, previous observations with the Australia Telescope Compact Array (ATCA) at 20 cm (1.3 GHz) and 6 cm (5 GHz) found a highly polarized radio sheath wrapping around the X-ray jet region, and the emission is very faint inside the sheath. In RCW 89, the radio emission is clumpy, and the peaks generally follow the X-ray knots (B. M. Gaensler et al. 1999, 2002). To better understand this highly unusual PWN/SNR system, high-resolution imaging is the key. Unfortunately, previous ATCA observations were taken before the CABB upgrade (W. E. Wilson et al. 2011) and they were optimized to detect a large-scale structure. In this study, we performed new observations with ATCA to obtain arcsecond-resolution images for a direct comparison with the Chandra data. The observations and analysis are described in Section 2. The results are shown in Section 3 and discussed in Section 4. In Section 5, we present the conclusion.

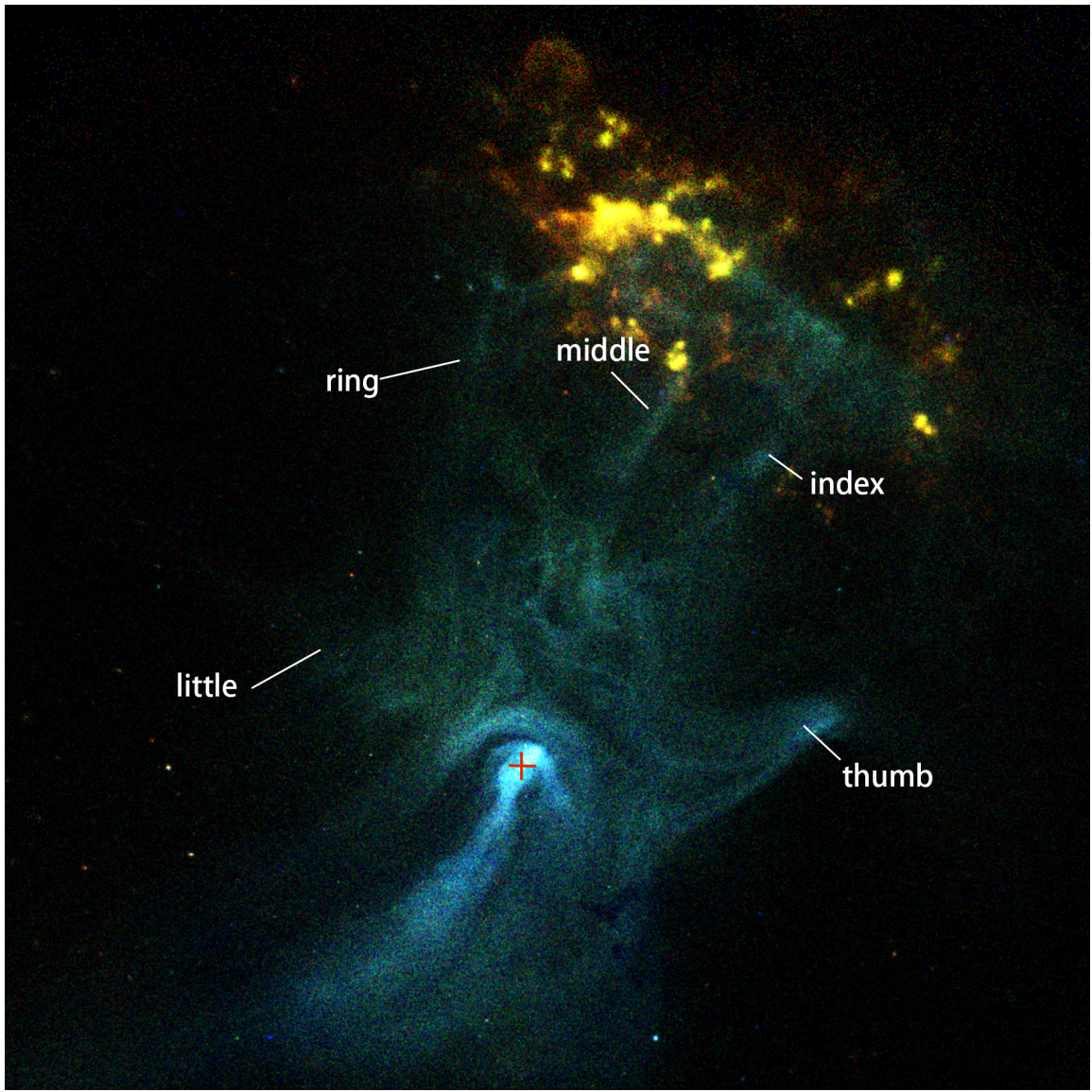

**Figure 1.** Chandra X-ray image of MSH 15−52 and RCW 89 in 0.5–7 keV (red), 1.2–2 keV (green), and 2–7 keV (blue) bands, made with all MSH 15−52 pointings taken between 2000 and 2020, exposure corrected and smoothed to 0″.6. PSR B1509−58 is marked by the red cross, and the X-ray fingers are indicated.

## 2. Observations and Data Analysis

Radio observations of MSH 15−52 were taken with ATCA on 5 days over 2011–2013, with on-source time of 8–11 hr each day. The 6 and 3 cm bands were observed simultaneously, centered at 5.5 and 9.0 GHz, respectively, with 2 GHz bandwidth at each band. Observations in 2011–2012 were made with compact array configurations from 367 m to 1.5 km and with mosaicking to cover the entire PWN and RCW 89, while the 2013 observations were made with the 6 km array configuration and single pointing centered on the pulsar position. The observation parameters are listed in Table 1. Combining all the data, the minimum baseline is 31 m, such that the 6 and 3 cm observations are sensitive to angular scales up to 6′.7 and 3′.3, respectively. PKS B1934−638 was observed for bandpass and flux density calibrations, and PKS 1520−58 was observed every 10 minutes for gain calibration.

We performed all data analysis using the MIRIAD package (R. J. Sault et al. 1995). After standard flagging and calibration procedures, we generate 6 and 3 cm images using the five data sets with the multifrequency synthesis technique. For Stokes $I$ image at 6 cm, we applied the Briggs's weighting scheme with the robust parameter of 0.4 to optimize between resolution and the level of the side lobe. Then, it is deconvolved with the maximum entropy method (MEM) by the task mosmem. We then applied self-calibration to further improve the signal-to-noise ratio (S/N). After cleaning, the map is restored with a beam of FWHM 2″.1 × 1″.8. The rms noise near the pointing center is ∼10 $\mu$Jy beam$^{-1}$, compatible with the theoretical sensitivity of 7.2 $\mu$Jy beam$^{-1}$. The 3 cm total intensity image is generated using the same process, but without self-calibration, as the emission is rather faint. The cleaned map is also restored with the same beam size as at 6 cm to allow a direct comparison. It has an rms noise of 12 $\mu$Jy beam$^{-1}$ near the center, close to the theoretical level of 9.4 $\mu$Jy beam$^{-1}$.

Table 1
ATCA Observations of MSH 15–52

| Observation Date | Array Configuration | Maximum Baseline (m) | Integration Time (hr) | No. of Pointings | Center Frequency (MHz) | Bandwidth (MHz) |
|---|---|---|---|---|---|---|
| 2011 Nov 14 | 750D | 4469 | 9.2 | 8 | 5500, 9000 | 2048 |
| 2011 Dec 3 | EW367 | 4408 | 8.5 | 8 | 5500, 9000 | 2048 |
| 2012 Feb 23 | 1.5D | 4439 | 9.1 | 8 | 5500, 9000 | 2048 |
| 2013 Feb 16 | 6A | 5939 | 11.1 | 1 | 5500, 9000 | 2048 |
| 2013 Mar 01 | 6B | 5969 | 9.4 | 1 | 5500, 9000 | 2048 |

For polarimetry, we generate Stokes $I$, $Q$, and $U$ images at 6 and 3 cm with slightly lower resolution to boost the S/N. We employ a Gaussian taper of 4″ FWHM and robust parameter of 0.4. The images at each band are then cleaned simultaneously with the MEM using the task pmosmem and then restored using a circular beam of 4″ FWHM. The final maps in both bands have an rms noise of ∼10 $\mu$Jy beam$^{-1}$. We generate polarization intensity and position angle (PA) maps using the task impol. The Ricean bias is corrected based on the rms noise levels above. We mask the 6 cm maps if either the polarized intensity has S/N < 5 or the total intensity has S/N < 11. Similarly, the 3 cm maps are masked if the polarized intensity has S/N < 3 or the total intensity has S/N < 7. We generate a rotation measure (RM) map of the field using lower-resolution (20″) PA maps at 3 and 6 cm. The typical uncertainty of the map is ∼20 rad m$^{-2}$. This is then used to correct for the Faraday rotation to deduce the intrinsic orientation of the magnetic field in MSH 15–52 and RCW 89. Our final PA maps have uncertainty ≲3°.

To measure the radio spectrum of various features in the PWN, we formed 6 and 3 cm total intensity images using the overlapping $u$–$v$ range, with identical imaging and deconvolution procedures as above. For a multiwavelength comparison, we also extracted the X-ray spectra of the same regions using a 45 ks Chandra observation taken in 2006 (ObsID 6117). X-ray spectra are fitted with an absorbed power law in the 0.5–7 keV range using the HEASARC XSPEC v12.12.0 software package (K. A. Arnaud 1996), with the absorption column density fixed at $N_H = 0.95 \times 10^{22}$ cm$^{-2}$.

## 3. Results

### 3.1. Morphology

The total intensity radio images of MSH 15–52 and RCW 89 at 6 and 3 cm are shown in Figures 2–6. The emission structure is very similar in both bands, albeit fainter at 3 cm, and the overall extent closely follows that of the X-ray emission. Our high-resolution maps reveal, for the first time, many detailed radio features of this system.

In the inner PWN, PSR B1509−58 is clearly detected in the total intensity images. We employed a Gaussian fit to obtain a position of R.A. = $15^h13^m55\overset{s}{.}6$, decl. = $-59°08'09\overset{''}{.}4$ (J2000.0) with flux densities of 0.24 mJy and 0.14 mJy at 6 and 3 cm, respectively. The position is in very good agreement with the X-ray position obtained with Chandra (B. M. Gaensler et al. 2002) and the recently reported radio timing position (M. J. Keith et al. 2024). Figure 3 shows a zoomed-in image of the inner PWN to illustrate the small-scale radio features near the pulsar. There is a linear filament of 1′ long and ∼10″ wide running through the pulsar from southeast to northwest at a PA of 115° (north to east), which does not align with the X-ray jet. We call it the "bar" and note that its width is resolved at the pulsar position, suggesting that it is likely a foreground or background object rather than a jet from the pulsar. The pulsar and the bar are wrapped by a sheath-shaped structure from the northwest, which was first reported by B. M. Gaensler et al. (1999). The sheath extends over 3′ to the southeast. Its emission is brightest and narrowest (∼20″ wide) in the northwest. It forms a semicircular arc of 0.8 diameter with a relatively sharp boundary. The northeastern part of the sheath appears to be composed of several filaments, and there is a hint of a spike southwest of the pulsar pointing north. It could be part of a filament running parallel to the semicircular arc, but deeper observation is needed to confirm this. South of the pulsar, the sheath becomes wider (∼1′) and more diffuse. Its inner boundary exhibits a few twists on both sides. The cavity inside the sheath is 40″ across at its narrowest, which is significantly wider than the X-ray jet (20″–30″).

The radio emission inside the sheath has very low emissivity. The faintest region is located at 0.5 south of the pulsar, with surface brightness below 20 $\mu$Jy beam$^{-1}$ (i.e., brightness temperature of 0.26 K) at 6 cm, close to the noise level of our observations. The brightness gradually increases toward the southeast and eventually becomes comparable to that of the surrounding sheath at ∼3′ from the pulsar. Beyond that, the cavity is no longer obvious, as the sheath blends into the surrounding diffuse emission. Figure 4 presents a comparison between radio and X-ray emissions of the inner PWN. The semicircular radio arc closely follows the outer arc in X-rays, and the radio emission fills the gap between the inner and outer X-ray arcs. However, there is no detection of the radio counterpart to all other small-scale X-ray features, including the compact blob northwest of the pulsar, the inner ring, the compact knots, the inner arc, and the large-scale jet.

North of the sheath, the radio emission shows a filamentary structure connecting to RCW 89 (see Figure 5). Both the thumb in the west and the little finger in the east are clearly seen in the 6 cm intensity map. The thumb in radio is ∼3.5 long and 1′ wide, pointing northwest. It well aligns with the X-ray counterpart and also shows brightness enhancement at the tip as in X-rays, although less significant. On the other hand, the little finger is much longer in radio than in X-rays. The inner part closely follows the X-ray emission and runs 1′ toward southeast. Beyond that, it turns northeast and extends ≳4′ with no X-ray counterpart. However, the other fingers are not obvious, as they overlap with the bright emission of RCW 89 (see Figure 6).

The middle finger shows X-ray emission of 1.8 length, extending from a bright knot in RCW 89 toward the southeast. In radio and H$\alpha$ (Q. A. Parker et al. 2005), there is also a linear feature at the same location, but it is significantly shorter (0.9) and has a slightly different orientation (see Figure 6). It is unclear if they are physically associated or just a chance alignment due to projection. We note that the index finger in the west also shows X-ray emission only, but no clear radio or H$\alpha$ emission is detected.

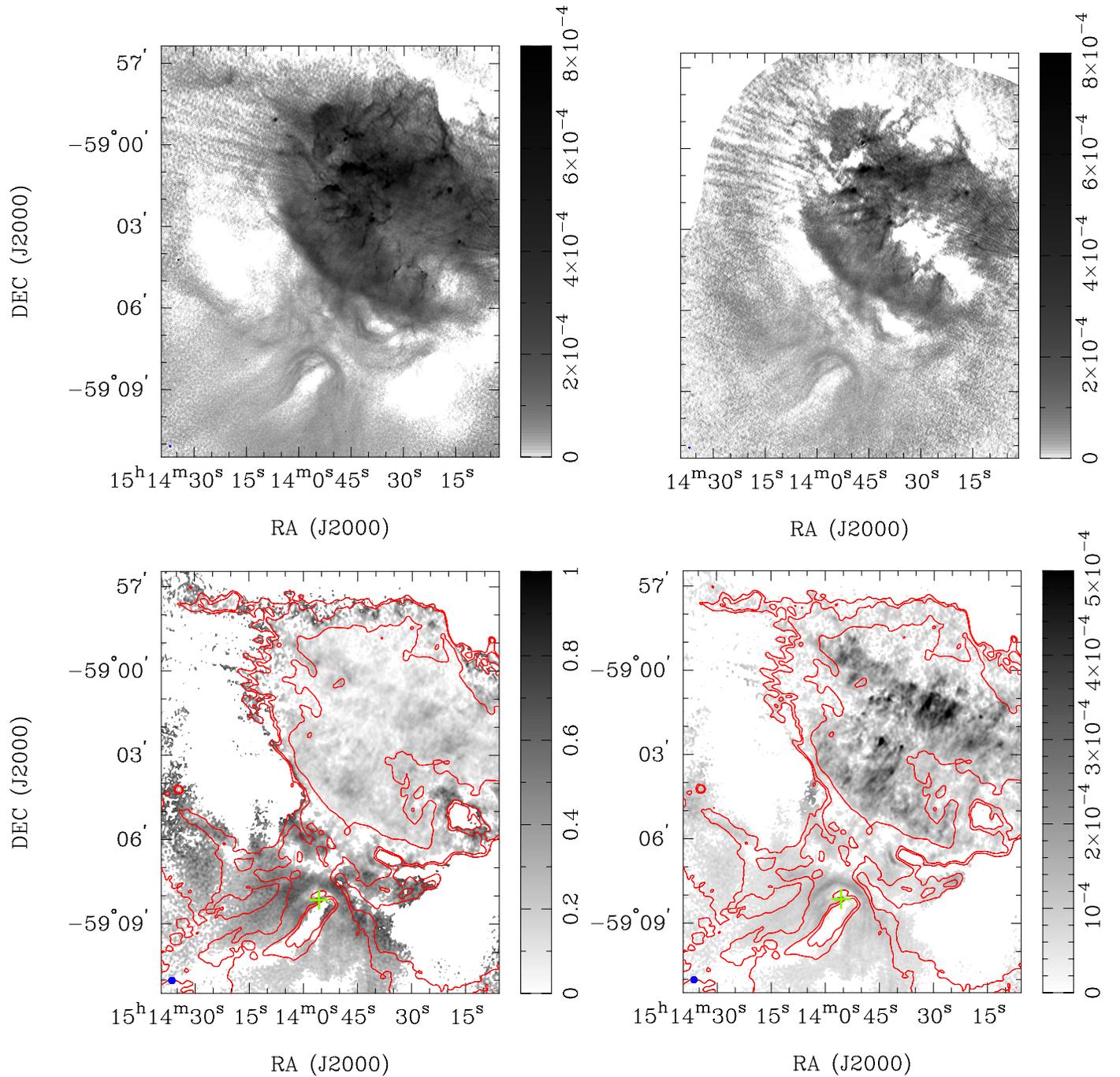

**Figure 2.** Top panels: total intensity radio maps of MSH 15−5*2* and RCW 89 at 6 cm (left) and 3 cm (right). Bottom panels: fractional polarization (left) and polarization intensity (right) maps at 6 cm. The beam size is shown by the ellipses in the lower left (the beams shown in the bottom panels indicate the beam of the contour). For the intensity maps, the grayscale bars have units of Jy beam$^{-1}$. The green cross marks the pulsar position. The contours are 6 cm total intensity levels at 0.3, 0.4, and 1.6 mK.

Figure 6 shows the radio intensity map of RCW 89 at 6 cm. The nebula has strong radio emission with brightness temperature up to ∼150 K near the center. The overall shape of the emission resembles a horseshoe of 10′ × 7′ opening toward the southwest. The rim shows a sharp boundary in the southeast, but it is more diffuse in other directions. The interior of the nebula consists of small-scale radio features, including bright knots in a complex network of filaments, all embedded in diffuse emission. In the same figure, we compare the radio map with X-ray and optical H$\alpha$ emission. The X-ray emission shows clumpy structure with bright knots and filaments located mainly at the center of the nebula. The knots are arranged in a horseshoe configuration similar to the overall extent of the radio emission but much smaller (∼3′ × 1.5′; see B. M. Gaensler et al. 2002; K. J. Borkowski et al. 2020), and many have radio counterparts. There are additional filaments in the outer part protruding radially outward (see also Figure 6). However, the blast wave in the middle part of RCW 89 identified in X-rays by K. J. Borkowski et al. (2020) has no radio counterpart. The optical emission, on the other hand, shows intertwined filaments across the entire nebula, and it does not coincide with the X-ray emission. Intriguingly, the radio structure shares many common features with both the X-ray and optical emissions. As Figure 6 indicates, most of the Hα filaments in the northern part of RCW 89 have clear radio counterparts, including most of the X-ray knots showing strong radio emission.

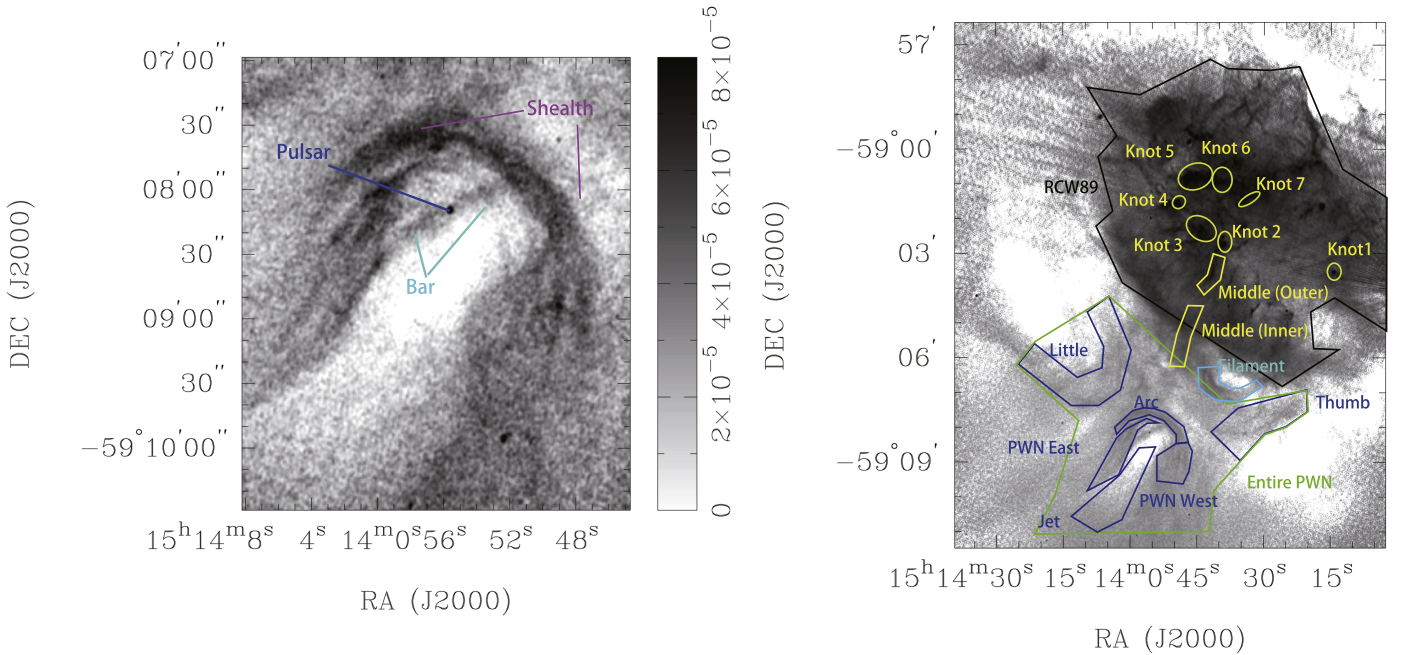

**Figure 3.** Left: zoom-in of the 6 cm intensity map to show the small-scale features in the inner PWN. Right: regions used for flux density measurements.

### 3.2. Polarization and Magnetic Field

The linear polarization fraction (PF) and polarization intensity maps of MSH 15−52 and RCW 89 at 6 cm are presented in Figures 2 and 4. We extract the polarized and total flux densities from selected regions shown in Figure 3 to estimate their PF. The results are listed in Table 2. The emission of MSH 15−52 is strongly linearly polarized, with PF up to ∼70% in the semicircular part of the sheath. The bar, the thumb, and the little finger also have a high PF of ∼40%. These values are higher than those in X-rays obtained from IXPE (20%–30%; R. W. Romani et al. 2023). Finally, for RCW 89, we find that the radio emission is also linearly polarized, but with a lower PF of ∼20%.

Figure 7 shows the foreground RM distribution. It varies across the field, from −700 rad m$^{-2}$ near the southeast end of the jet to +500 rad m$^{-2}$ in the northwestern edge of RCW 89. Such a large variation could be contributed by the dense cloud in the surroundings as discussed in Section 4 below. The RM in the inner PWN is ∼250 rad m$^{-2}$, broadly consistent with the pulsar value of 223 rad m$^{-2}$ (M. J. Keith et al. 2024).

After correcting for the foreground RM, the intrinsic magnetic field orientation of MSH 15−52 and RCW 89 is shown in Figures 2–6. The B field is highly ordered, and it generally follows the nebular structure. In the inner PWN, the field lines well align with the sheath, particularly the semicircular arc in the northwest and the filament that runs across the pulsar (see Figure 4). They also closely follow the structure thumb and the little finger (see Figure 5). North of the sheath, the B field points northwest toward RCW 89, consistent with the orientation of the three inner fingers in X-rays (see Figure 6). All these well agree with the previous radio result from ATCA (B. M. Gaensler et al. 1999) and the recent X-ray polarimetry results from the IXPE (R. W. Romani et al. 2023). Inside RCW 89, the field is mainly along the southeast–northwest direction, and it turns slightly northward beyond the opening of the horseshoe structure. This confirms the finding using lower-resolution single-dish observations at the same bands (D. K. Milne et al. 1993). We note that the B-field direction shows no small-scale variation within RCW 89 and has only very weak correlation with the radio filament structure. In the northwest beyond the X-ray emission, our observations lack sensitivity to detect any radio polarization signal. We are therefore unable to map the B-field orientation in that region.

### 3.3. Spectrum

The 3 and 6 cm radio flux density measurements of selected regions in MSH 15−52 and RCW 89 are listed in Table 2, and the regions are shown in Figure 3. While simple background subtraction has been performed, we caution that these measurements could be subject to large (≳20%) uncertainties due to side lobes and a highly complicated background. For the same reason, the spectral index reported in the table, which is calculated based on the 6 and 3 cm flux densities, could also have large uncertainties and should be taken as indicative only. For PSR B1509−58, we combine our data with the Parkes measurement at 1.4 GHz (1.43 mJy; F. Jankowski et al. 2018) to obtain a spectral index[5] $\alpha = -1.4$, which is consistent with other radio pulsars (mean $\alpha = -1.6$ with standard deviation of 0.54; F. Jankowski et al. 2018).

---
[5] The spectral index $\alpha$ is defined as $S_\nu \propto \nu^\alpha$.

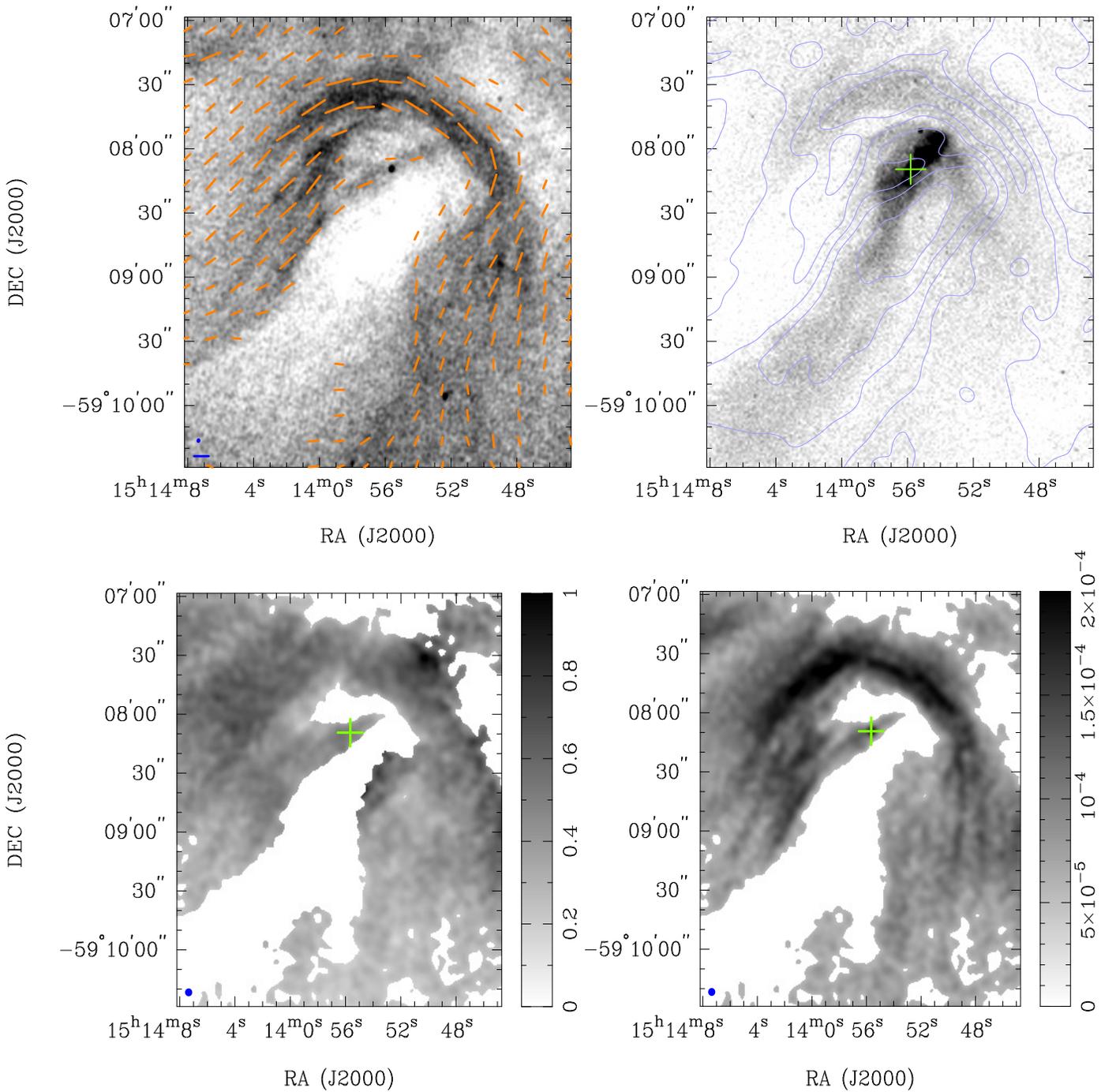

**Figure 4.** Top left: 6 cm radio intensity map of the inner region of MSH 15–52, overlaid with the intrinsic magnetic field vectors. The vector lengths are proportional to the polarized intensity at 6 cm. The scale bar at the lower left represents 0.1 mJy beam$^{-1}$. Top right: Chandra X-ray image of the same field in 0.5–7 keV range, overlaid with the 6 cm radio total intensity contours at levels of 0.12, 0.24, …, 0.60 mK. Bottom panel: fractional polarization (left) and polarization intensity (right) maps at 6 cm. The beam size of all radio maps is shown at the lower left.

Table 2 indicates that most of the PWN features and knots in RCW 89 have flat radio spectra with index $\alpha \gtrsim -1.0$. However, there is no clear pattern across the PWN, which could be due to the complex background. In contrast, the X-ray spectrum steepens as it moves away from the pulsar, confirming the results reported in previous studies (e.g., H. An et al. 2014). We note that the knots in RCW 89 are known to show thermal emission (see Y. Yatsu et al. 2005; K. J. Borkowski et al. 2020), and fitting with a simple power law results in large photon indexes ($\Gamma > 5$).

In Figure 8, we plot the multiwavelength spectra of the entire MSH 15–52 and selected small-scale features, including the jet, the arc, and the thumb. The latter two show corresponding structure in both radio and X-ray, while the jet has no radio detection, so only upper limits are plotted. It is clear that in all these cases, extrapolation of the X-ray spectra is incompatible with the radio flux points, implying that the spectra in these two bands cannot be joined with a simple power law.

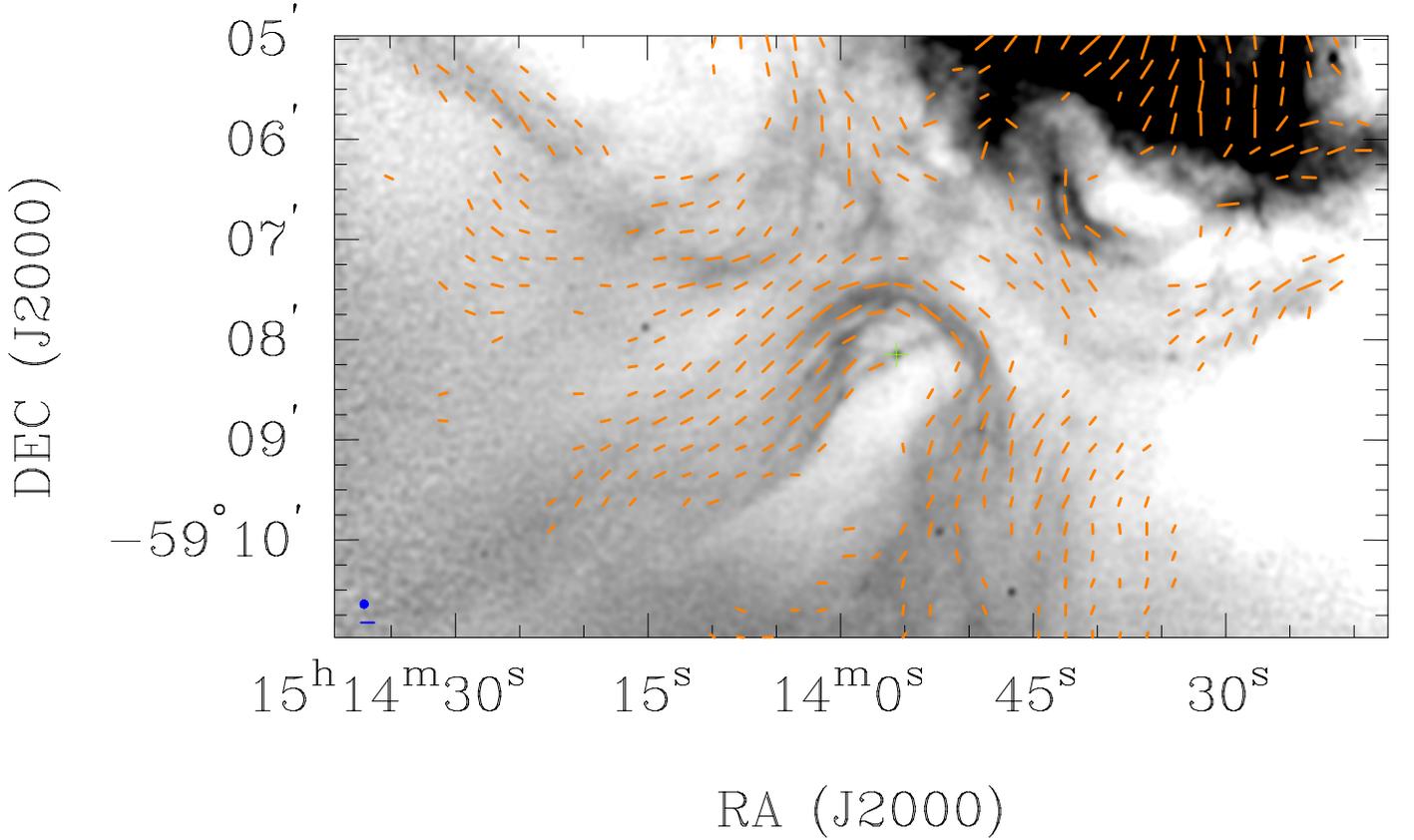

**Figure 5.** 6 cm total intensity map smoothed to 4″ resolution to show the radio filaments in the X-ray finger region. The vectors indicate the intrinsic $B$-field orientation and their lengths are proportional to the polarization intensity at 6 cm. The beam size is shown on the lower left, and the scale bar below it corresponds to the polarized intensity of 0.1 mJy beam$^{-1}$.

**Table 2**
Radio and X-Ray Flux Measurements and Spectra of Different Regions in MSH 15–52 and RCW 89

| Region | $F_{5.5\,\mathrm{GHz}}$ (mJy) | $F_{9\,\mathrm{GHz}}$ (mJy) | $\alpha$ | PF | $\Gamma$ | $F_X^{\mathrm{unabs}}$ ($10^{-12}$ erg cm$^{-2}$ s$^{-1}$) |
|---|---|---|---|---|---|---|
| Pulsar | 0.2 ± 0.01 | 0.11 ± 0.01 | −1.2 ± 0.3 | ... | 0.4 | 3.2 |
| Bar | 2 ± 0.1 | 1.5 ± 0.1 | −0.6 ± 0.3 | 0.45 ± 0.05 | 1.5 | 3.0 |
| Arc | 20 ± 1 | 14 ± 1 | −0.7 ± 0.3 | 0.70 ± 0.04 | 1.9 | 3.1 |
| PWN East | 19 ± 1 | 11 ± 0.9 | −1.1 ± 0.2 | 0.53 ± 0.03 | (1.8) | (1.8) |
| PWN West | 25 ± 1 | 17 ± 1.1 | −0.8 ± 0.2 | 0.44 ± 0.03 | (1.7) | (3.2) |
| Jet | (<19) | (<17) | ... | ... | 1.8 | 13 |
| Filament | 27 ± 1.2 | 13 ± 1 | −1.5 ± 0.2 | 0.35 ± 0.04 | 2.1 | 2.9 |
| Thumb | 54 ± 4 | 44 ± 3 | −0.4 ± 0.3 | 0.43 ± 0.05 | 2.0 | 4.5 |
| Middle finger (inner) | (49 ± 4) | (34 ± 3) | (−0.7 ± 0.3) | 0.25 ± 0.03 | 2.3 | 7.9 |
| Middle finger (outer) | (51 ± 4) | (39 ± 3) | (−0.5 ± 0.4) | 0.11 ± 0.04 | 2.2 | 1.6 |
| Little finger | 76 ± 7 | 36 ± 5 | −1.5 ± 0.2 | 0.29 ± 0.05 | 2.3 | 8.0 |
| Entire PWN | 697 ± 38 | 438 ± 31 | −0.9 ± 0.3 | 0.30 ± 0.02 | 1.8 | 90 |
| Knot 1 | 22 ± 0.6 | 15 ± 0.4 | −0.8 ± 0.1 | 0.08 ± 0.01 | 6.0 | 7.7 |
| Knot 2 | 37 ± 0.5 | 27 ± 0.5 | −0.6 ± 0.1 | 0.14 ± 0.01 | 5.4 | 9.7 |
| Knot 3 | 77 ± 2 | 60 ± 2 | −0.5 ± 0.1 | 0.13 ± 0.01 | 5.2 | 6.0 |
| Knot 4 | 32 ± 1 | 22 ± 1 | −0.8 ± 0.1 | 0.04 ± 0.01 | 5.6 | 13 |
| Knot 5 | 140 ± 4 | 98 ± 2 | −0.7 ± 0.1 | 0.06 ± 0.01 | 6.3 | 85 |
| Knot 6 | 93 ± 4 | 63 ± 3 | −0.8 ± 0.2 | 0.10 ± 0.01 | 6.0 | 24 |
| Knot 7 | 45 ± 2 | 29 ± 1 | −0.9 ± 0.1 | 0.20 ± 0.01 | 5.4 | 11 |
| RCW 89 | 5300 ± 242 | 3000 ± 188 | −1.2 ± 0.2 | 0.16 ± 0.04 | ... | ... |

**Note.** $F_{5.5\,\mathrm{GHz}}$ and $F_{9\,\mathrm{GHz}}$ are the radio flux densities at 5.5 GHz and 9 GHz, respectively, and $\alpha$ is the radio spectral index directly calculated from these two values. The polarization fraction (PF) is measured from the 6 cm (i.e., 5.5 GHz) images. The X-ray photon index $\Gamma$ and unabsorbed flux $F_X^{\mathrm{unabs}}$ are in the 0.5–7 keV energy range, obtained from a power-law fit with column density fixed at $N_{\mathrm{H}} = 9.5 \times 10^{21}$ cm$^{-2}$. Values in parentheses denote that there is no clear corresponding radio or X-ray structure; we nonetheless report the flux density or spectral measurement in the region. Upper limits for the radio flux densities in the jet region are at $3\sigma$ confidence level.

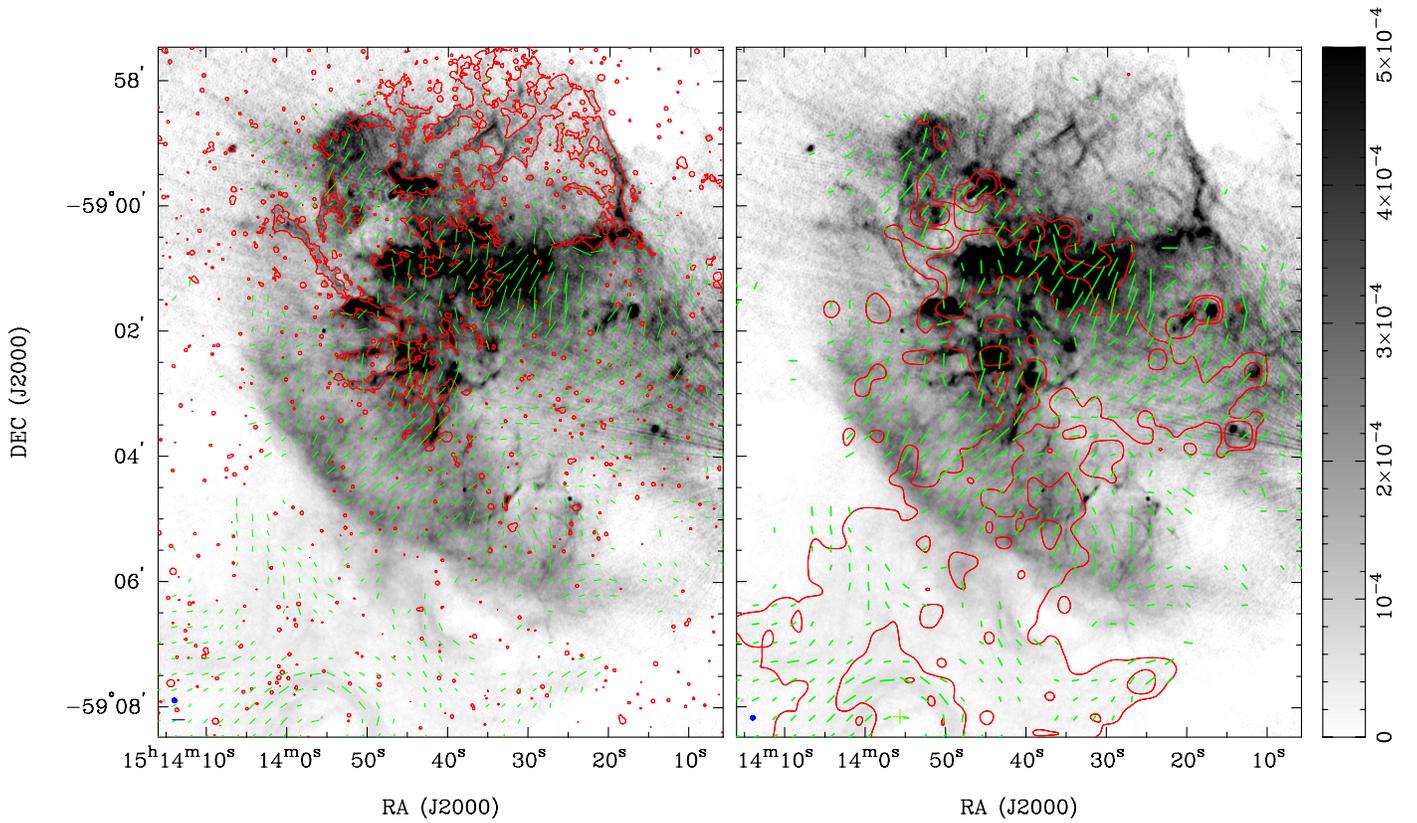

**Figure 6.** Radio total intensity map of RCW 89 at 6 cm, overlaid with Hα contours (left) and X-ray contours (right). The H-alpha intensity is from the SuperCOSMOS H-alpha Survey (Q. A. Parker et al. 2005), with bright stars removed. The grayscale is in units of Jy beam$^{-1}$. The vectors indicate the intrinsic direction of the magnetic field with the length corresponding to the polarization intensity at 6 cm. The blue scale bar at the lower left represents 0.1 mJy beam$^{-1}$, and the blue ellipse shows the beam size of $2\rlap{.}''1 \times 1\rlap{.}''8$ for the radio intensity maps.

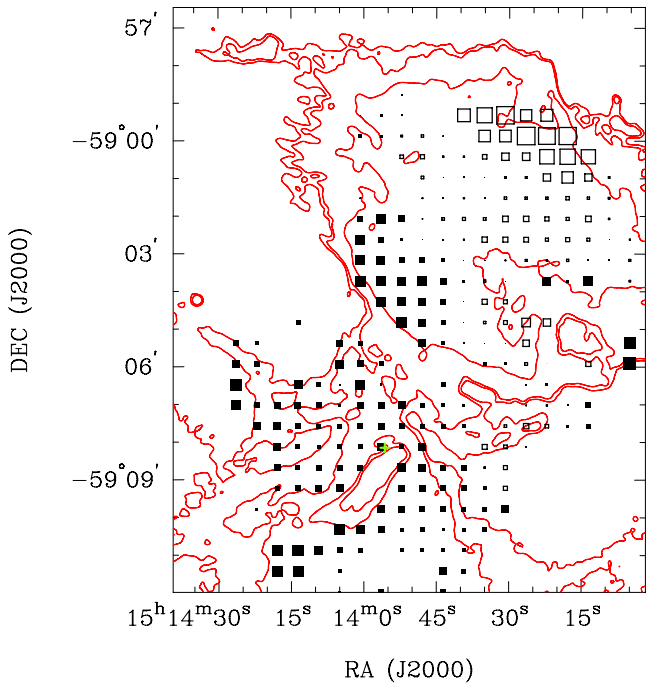

**Figure 7.** RM toward MSH 15–52 and RCW 89, smoothed to a resolution of $30''$. The box size represents the RM values, ranging from $-700$ rad m$^{-2}$ to $+500$ rad m$^{-2}$ in the field, with the open and solid squares indicating negative and positive values, respectively. The contours are total intensity levels at 0.3, 0.4, and 1.6 mK, and the green cross marks the position of PSR B1509−58. The typical uncertainty of the RM measurement is 10 rad m$^{-2}$.

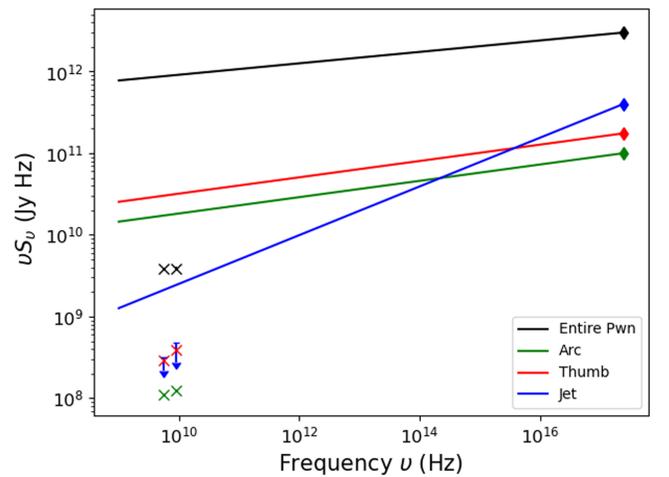

**Figure 8.** Broadband spectra of the overall PWN of MSH 15–52 and small-scale features. The lines show the best-fit X-ray spectrum at 1 keV extrapolated to 6 cm wavelength. The crosses on the left are obtained with ATCA observation at 6 and 3 cm observations, and the upper limits (blue arrows) show $3\sigma$ rms noise level of 6 and 3 cm observations.

## 4. Discussion

### 4.1. Structure of MSH 15–52

Our high-resolution radio images of MSH 15–52 at 6 and 3 cm reveal the detailed structure of the PWN. Its overall morphology consists of a sheath in the south and protrusions in the north, confirming the previous findings from lower-resolution data (B. M. Gaensler et al. 1999, 2002). Our new images discover that the radio nebula mainly consists of filaments, which are highly linearly polarized and well aligned with the magnetic field orientation, indicating synchrotron nature. Similar filamentary structure is commonly found in young PWNe, such as the Crab, G21.5−0.9, G54.1+0.3, and 3C 58 (see S. Safi-Harb et al. 2001; M. F. Bietenholz 2006; C. C. Lang et al. 2010; T. Temim et al. 2024). On the other hand, what makes MSH 15−52 unique is the peculiar sheathlike radio feature surrounding the pulsar. Since discovered over two decades ago, this remains the only known example, and its nature is still difficult to interpret.

We found that the northwestern part of the sheath, i.e., the semicircular arc, well matches the outer arc in X-rays. If this represents an equatorial torus as those commonly seen in young X-ray PWNe, then it will be a rare case that shows a radio counterpart, in contrast to the Crab, Vela, B1706−44, and G54.1+0.3 (see, e.g., R. Dodson et al. 2003; C. C. Lang et al. 2010; Y. H. Liu et al. 2023). B. M. Gaensler et al. (2002) argued based on the synchrotron cooling timescale that the arc cannot be the torus but could be a wisp as those seen in the Crab Nebula in radio, optical, and X-ray bands. Such wisps are bright, narrow, ring-shaped emissions in the equatorial plane (M. F. Bietenholz et al. 2004; T. Schweizer et al. 2013), and they could be resulting from the combined effect of Doppler boosting and magnetic field enhancement by turbulence in the flow (e.g., B. Olmi et al. 2015). In our case, the radio arc of MSH 15−52 is much thicker than the wisps in the Crab and 3C 58 ($20''\approx 0.5$ pc at 5 kpc versus $\sim 0.02$ pc; D. A. Frail & D. A. Moffett 1993; G. Dubner et al. 2017). Moreover, the radio wisp of the Crab is found to be moving outward at an apparent speed of $\sim 0.3c$ (M. F. Bietenholz et al. 2004), but the overall structure of the sheath here appears to be stable in the radio and X-ray images spanning over two decades (see also T. DeLaney et al. 2006). As the light-crossing time for the 0.5 pc width arc is only 1.6 yr, this poses an upper limit of $<0.01c$ for any motion. Moreover, the wisps in both in 3C 58 and the Crab have brightness profiles much more peaked near the PWN symmetric axis, which is very different from what we see in the radio arc here.

Alternatively, the sheath could be a synchrotron filament with similar nature as those in the northern part of the PWN, particularly those between the thumb and the little finger. These filaments are commonly seen in high-resolution radio imaging of young PWNe inside SNRs. They represent local enhancement of the $B$ field or the particle density. Various formation mechanisms have been proposed, including interaction between pulsar wind and thermal filaments (see S. P. Reynolds 1988), Rayleigh–Taylor instabilities when the pulsar wind expands into the surrounding SNR ejecta (N. Bucciantini et al. 2004), and magnetic structure torn from the toroidal field due to kink instabilities (P. Slane et al. 2004). In addition, we suggest that the filaments could also be formed when the pulsar wind injects into existing magnetic flux tubes. The latter could possibly be due to turbulence or compression by the supernova shock. To test these ideas, further simulation works are needed to provide a better understanding of the complex interplay between the pulsar wind and the supernova ejecta.

In all the interpretations above (either torus, wisp, or filament), if the radio arc physically lies in the pulsar equatorial plane, it remains to be explained why it forms only a partial ring and opens up in the southeast. This would require a highly unusual flow structure, possibly shaped by anisotropic ambient pressure. Such an effect is commonly observed in evolved PWN systems, either due to the passage of supernova reverse shock or supersonic pulsar motion. Both cases could lead to highly elongated, comet-like nebular structure surrounding the pulsar with bright X-ray and radio emissions near the head (see, e.g., B. M. Gaensler et al. 2004; T. Temim et al. 2009). Here, we focus on the reverse-shock interaction scenario, since the pulsar motion is unlikely to be supersonic (see B. M. Gaensler et al. 2002). It was proposed that the supernova blast wave could have first encountered dense interstellar materials in the northwest (M. Tsirou et al. 2017), such as the H I filament G. M. Dubner et al. (2002) reported. As a result, the reverse shock would have arrived at the PWN first from that direction, sweeping the pulsar wind to the southeast to form an elongated relic nebula. This can explain the overall morphology of the sheath. As shown in the previous radio study of the Snail (G327.1−1.1), a relic PWN could show highly linearly polarized synchrotron filaments similar to those seen here (Y. K. Ma et al. 2016). However, one major issue of this picture is that hydrodynamic simulations generally do not expect the PWN structure pointing opposite to the shock propagation direction (e.g., T. Temim et al. 2015, 2017; C. Kolb et al. 2017; P. Slane et al. 2018). It is therefore nontrivial to explain the fingers and filament structure in the north toward RCW 89 (see discussion in Section 4.2 below). Further studies are needed to determine whether these structures can survive or be quickly regenerated after the shock passage.

Another puzzling feature of MSH 15–52 is the bright X-ray jet inside the sheath that shows no radio counterpart (see Figure 4). Similar anticorrelation is observed in the inner regions of young PWNe, including Vela (R. Dodson et al. 2003) and the PWN of PSR B1706−44 (Y. H. Liu et al. 2023), and at the tip of the bow-shock PWN of PSR J1509−5850 (C. Y. Ng et al. 2010). As Figure 8 clearly indicates, the broadband spectrum of the jet from radio to X-rays is incompatible with a single power law. Given that the X-ray emitting particles in the jet are uncooled (B. M. Gaensler et al. 2002), the result implies an intrinsic low-energy cutoff in the injected particle distribution, and the X-ray- and radio-emitting particles could have distinct origins (see C. C. Lang et al. 2010). This also applies to other small-scale features in the PWN, including the arc and the thumb (see Figure 8). For the arc, the radio emission shows much higher PF than the X-ray counterpart. In particular, its radio PF $\approx 0.7$ is close to the synchrotron limit, implying a very low level of turbulence (see R. Bandiera & O. Petruk 2016). This adds support to the above idea of a different origins.

Recently, jetlike features have been observed in several bow-shock PWNe. They are detected in X-ray only and are highly collimated over a few parsecs (see J. T. Dinsmore & R. W. Romani 2024). They are suggested to be due to high-energy particles leaked out from the shock that gyrate along magnetic field lines in the interstellar medium (R. Bandiera 2008). A similar mechanism may help explain the nature of the X-ray jet here in MSH 15−52. This could also be applied to the three inner fingers (i.e., the index, middle, and ring fingers) in the north, which are proposed to be the counterjets (B. M. Gaensler et al. 2002; Y. Yatsu et al. 2005). Finally, we note that if the reverse-shock interaction scenario holds, then these fingers share some similarities with the prong-like X-ray structure found in G327.1−1.1 (T. Temim et al. 2009, 2015). They both show no radio emission and point against the reverse-shock propagation direction. The only major difference is that the fingers in our case are slightly longer (15 pc versus 10 pc) and are not directly connected to the inner PWN.

### 4.2. Nature of RCW 89

Our detection of linearly polarized emission from RCW 89 clearly indicates that this is synchrotron radiation from relativistic particles. If the supernova explosion site is near the current pulsar position, the polarization direction then implies a predominately radial $B$ field in the remnant, as typically found in young SNRs (see E. M. Reynoso et al. 1997; X. H. Sun et al. 2011; W. D. Cotton et al. 2024). The overall PF of ∼13% is also in line with other young remnants (see G. Dubner & E. Giacani 2015). However, RCW 89 is not a typical shell-type SNR. The northwestern edge of the X-ray remnant shows nonthermal spectrum and is found to be moving outward at 4000 km s$^{-1}$. It is therefore suggested to be a blast wave (K. J. Borkowski et al. 2020). Intriguingly, we found no corresponding radio structure at that position, and the radio remnant extends well beyond (>1′.5) the X-ray edge (see Figure 6). This is in contrast to the coincident radio and X-ray boundaries generally seen in shell-type SNRs (V. Domcek et al. 2017). The offset here is not easy to understand, unless it is due to projection effect or the X-ray blast wave is associated with the reverse shock. Similar offset is more often found in mixed-morphology SNRs (MMSNRs), which are remnants interacting with surrounding dense clouds (see J. Vink 2012). In our case, the supernova shock of RCW 89 is suggested to have recently encountered a dense wall in the northwest, based on its overall flattened shape (B. M. Gaensler et al. 1999). This is supported by the discovery of a dense (12 cm$^{-3}$) H I filament in that direction (G. M. Dubner et al. 2002). Moreover, the coexistence of radio, X-ray, and Hα filaments (see Figure 6) gives additional evidence that the shock wave is expanding into cool and dense medium (see M. Stupar et al. 2007). That being said, we note that MMSNRs show only thermal X-ray emission (except G346.6−0.2; K. Auchettl et al. 2017), and it has no coincidence with the radio filaments (e.g., R. L. Shelton et al. 2004; G. Castelletti et al. 2007; E. Giacani et al. 2009), which is different from what we find here.

While the thermal X-ray knots inside RCW 89 are suggested to be hot spots in the H I cloud lit up by a pulsar counterjet (B. M. Gaensler et al. 1999; G. M. Dubner et al. 2002; Y. Yatsu et al. 2009), we find no evidence of this in the high-resolution radio and X-ray images. It has been recently reported that the knots are composed of heavy elements and are moving out at a high speed (K. J. Borkowski et al. 2020). They are therefore more likely to be supernova ejecta knots similar to those found in Cas A. Their interaction with the H I cloud not only generates thermal X-rays but also can accelerate particles. This could give rise to the nonthermal radio clumps at the knot positions (see Figure 6). Finally, we mention that the knots generally align with the inner X-ray finger pointing direction, but further theoretical or simulation works are needed to investigate their physical connection, e.g., whether the fast-moving knots can leave behind low-density channels for the pulsar wind to fill in, or if there is enhanced magnetic field trailing behind their motion to generate the fingerlike features.

### 5. Conclusion

In this paper, we present a high-resolution radio study of MSH 15–52 and RCW 89 using ATCA observations at 3 and 6 cm. The new images reveal the details of the small-scale features in the PWN. We found a complex network of filaments in the PWN. They are highly linearly polarized, and the intrinsic magnetic field orientation well follows the filamentary structure. This could be resulted from the interaction between the pulsar wind and the surrounding supernova ejecta. In particular, the jet and the sheath features could possibly be formed by a gradient in the ambient pressure, although the physical origin of such a gradient is unclear.

A detailed multiwavelength comparison shows a significant difference between the radio and X-ray structures of the PWN. Some prominent X-ray features, including the one-sided jet in the south and the inner three fingers in the north, are not detected in the radio band. This implies a low-energy cutoff in the particle distribution. We suggest that some of these structures could have a similar nature as misaligned jets found in bow-shock PWNe, which are due to high-energy particles leaked out from the shock.

For RCW 89, our result show that its structure is largely distinct from typical young shell-type SNRs. The radio emission is patchy and exhibits a good correlation with the Hα and X-ray knots and filaments. It also extends significantly beyond the X-ray emission, similar to the case of MMSNRs. All these provide support to the picture that RCW 89 is interacting with a dense H I cloud in the surroundings. Finally, we note that the sharp boundary of the X-ray emission is suggested to be the supernova blast wave, but no corresponding radio counterpart is found, which remains to be explained.

In summary, MSH 15–52 and RCW 89 is a remarkable system that shows many unique features not found in other young sources. There are, however, still many open questions regarding the formation and evolution of these structures. Further simulation works are needed to provide better understanding of the complex interplay between the pulsar wind and the supernova ejecta.

### Acknowledgments

We thank Lisa Harvey-Smith for help carrying out the ATCA observations in 2012–2013. The Australia Telescope Compact Array is part of the Australia Telescope National Facility (grid.421683.a) which is funded by the Australian Government for operation as a National Facility managed by CSIRO. We thank the *ApJ* referee for all the suggestions. C.-Y. N. and S.Z. are supported by a GRF grant of the Hong Kong Government under HKU 17304524. N.B. was supported by the INAF MiniGrant "PWNnumpol—Numerical Studies of Pulsar Wind Nebulae in The Light of IXPE."

This paper employs a list of Chandra data sets, obtained by the Chandra X-ray Observatory, contained in the Chandra Data Collection (CDC) cdc.419 doi:10.25574/cdc.419.

### ORCID iDs

S. Zhang  https://orcid.org/0000-0002-2096-6051
C.-Y. Ng  https://orcid.org/0000-0002-5847-2612
N. Bucciantini  https://orcid.org/0000-0002-8848-1392

**References**

Aharonian, F., Akhperjanian, A. G., Aye, K. M., et al. 2005, A&A, 435, L17
An, H., Madsen, K. K., Reynolds, S. P., et al. 2014, ApJ, 793, 90
Arnaud, K. A. 1996, in ASP Conf. Ser. 101, Astronomical Data Analysis Software and Systems V, ed. G. H. Jacoby & J. Barnes (San Francisco, CA: ASP), 17
Auchettl, K., Ng, C. Y., Wong, B. T. T., Lopez, L., & Slane, P. 2017, ApJ, 847, 121
Bandiera, R. 2008, A&A, 490, L3
Bandiera, R., & Petruk, O. 2016, MNRAS, 459, 178
Bietenholz, M. F. 2006, ApJ, 645, 1180
Bietenholz, M. F., Hester, J. J., Frail, D. A., & Bartel, N. 2004, ApJ, 615, 794
Borkowski, K. J., Reynolds, S. P., & Miltich, W. 2020, ApJL, 895, L32
Brazier, K. T. S., & Becker, W. 1997, MNRAS, 284, 335
Bucciantini, N., Amato, E., Bandiera, R., Blondin, J. M., & Del Zanna, L. 2004, A&A, 423, 253
Castelletti, G., Dubner, G., Brogan, C., & Kassim, N. E. 2007, A&A, 471, 537
Cotton, W. D., Kothes, R., Camilo, F., et al. 2024, ApJS, 270, 21
Cusumano, G., Mineo, T., Massaro, E., et al. 2001, A&A, 375, 397
DeLaney, T., Gaensler, B. M., Arons, J., & Pivovaroff, M. J. 2006, ApJ, 640, 929
Dinsmore, J. T., & Romani, R. W. 2024, ApJ, 976, 4
Dodson, R., Lewis, D., McConnell, D., & Deshpande, A. A. 2003, MNRAS, 343, 116
Domcek, V., Vink, J., Arias, M., & Zhou, P. 2017, in The X-ray Universe 2017, ed. J. U. Ness & S. Migliari (Rome, Italy) 264
Dubner, G., Castelletti, G., Kargaltsev, O., et al. 2017, ApJ, 840, 82
Dubner, G., & Giacani, E. 2015, A&ARv, 23, 3
Dubner, G. M., Gaensler, B. M., Giacani, E. B., Goss, W. M., & Green, A. J. 2002, AJ, 123, 337
Forot, M., Hermsen, W., Renaud, M., et al. 2006, ApJL, 651, L45
Frail, D. A., & Moffett, D. A. 1993, ApJ, 408, 637
Gaensler, B. M., Arons, J., Kaspi, V. M., et al. 2002, ApJ, 569, 878
Gaensler, B. M., Brazier, K. T. S., Manchester, R. N., Johnston, S., & Green, A. J. 1999, MNRAS, 305, 724
Gaensler, B. M., & Slane, P. O. 2006, ARA&A, 44, 17
Gaensler, B. M., van der Swaluw, E., Camilo, F., et al. 2004, ApJ, 616, 383
Giacani, E., Smith, M. J. S., Dubner, G., et al. 2009, A&A, 507, 841
Greiveldinger, C., Caucino, S., Massaglia, S., Oegelman, H., & Trussoni, E. 1995, ApJ, 454, 855
Jankowski, F., van Straten, W., Keane, E. F., et al. 2018, MNRAS, 473, 4436
Keith, M. J., Johnston, S., Karastergiou, A., et al. 2024, MNRAS, 530, 1581
Kesteven, M. J. L. 1968, AuJPh, 21, 739
Kolb, C., Blondin, J., Slane, P., & Temim, T. 2017, ApJ, 844, 1
Lang, C. C., Wang, Q. D., Lu, F., & Clubb, K. I. 2010, ApJ, 709, 1125
Liu, Y. H., Ng, C. Y., & Dodson, R. 2023, ApJ, 945, 82
Ma, Y. K., Ng, C. Y., Bucciantini, N., et al. 2016, ApJ, 820, 100
Marsden, D., Blanco, P. R., Gruber, D. E., et al. 1997, ApJL, 491, L39
Mills, B. Y., Slee, O. B., & Hill, E. R. 1961, AuJPh, 14, 497
Milne, D. K., Caswell, J. L., & Haynes, R. F. 1993, MNRAS, 264, 853
Ng, C. Y., Gaensler, B. M., Chatterjee, S., & Johnston, S. 2010, ApJ, 712, 596
Ng, C. Y., & Romani, R. W. 2004, ApJ, 601, 479
Ng, C. Y., & Romani, R. W. 2008, ApJ, 673, 411
Olmi, B., Del Zanna, L., Amato, E., & Bucciantini, N. 2015, MNRAS, 449, 3149
Parker, Q. A., Phillipps, S., Pierce, M. J., et al. 2005, MNRAS, 362, 689
Reynolds, S. P. 1988, ApJ, 327, 853
Reynoso, E. M., Moffett, D. A., Goss, W. M., et al. 1997, ApJ, 491, 816
Rodgers, A. W., Campbell, C. T., & Whiteoak, J. B. 1960, MNRAS, 121, 103
Romani, R. W., Wong, J., Di Lalla, N., et al. 2023, ApJ, 957, 23
Safi-Harb, S., Harrus, I. M., Petre, R., et al. 2001, ApJ, 561, 308
Sako, T., Matsubara, Y., Muraki, Y., et al. 2000, ApJ, 537, 422
Sault, R. J., Teuben, P. J., & Wright, M. C. H. 1995, in ASP Conf. Ser. 77, Astronomical Data Analysis Software and Systems IV, ed. R. A. Shaw, H. E. Payne, & J. J. E. Hayes (San Francisco, CA: ASP), 433
Schweizer, T., Bucciantini, N., Idec, W., et al. 2013, MNRAS, 433, 3325
Seward, F. D., & Harnden, F. R., Jr. 1982, ApJL, 256, L45
Seward, F. D., Harnden, F. R., Jr., & Clark, D. H. 1983, ApJ, 267, 698
Shelton, R. L., Kuntz, K. D., & Petre, R. 2004, ApJ, 615, 275
Slane, P., Helfand, D. J., van der Swaluw, E., & Murray, S. S. 2004, ApJ, 616, 403
Slane, P., Lovchinsky, I., Kolb, C., et al. 2018, ApJ, 865, 86
Stupar, M., Parker, Q. A., Filipović, M. D., et al. 2007, MNRAS, 381, 377
Sun, X. H., Reich, W., Wang, C., Han, J. L., & Reich, P. 2011, A&A, 535, A64
Temim, T., Laming, J. M., Kavanagh, P. J., et al. 2024, ApJL, 968, L18
Temim, T., Slane, P., Gaensler, B. M., Hughes, J. P., & Van Der Swaluw, E. 2009, ApJ, 691, 895
Temim, T., Slane, P., Kolb, C., et al. 2015, ApJ, 808, 100
Temim, T., Slane, P., Plucinsky, P. P., et al. 2017, ApJ, 851, 128
Tsirou, M., Gallant, Y., Zanin, R., Terrier, R. & H. E. S. S. Collaboration 2017, 35th IICRC (Busan), 301, 681
Vink, J. 2012, A&ARv, 20, 49
Wilson, W. E., Ferris, R. H., Axtens, P., et al. 2011, MNRAS, 416, 832
Yatsu, Y., Kawai, N., Kataoka, J., et al. 2005, ApJ, 631, 312
Yatsu, Y., Kawai, N., Shibata, S., & Brinkmann, W. 2009, PASJ, 61, 129